\magnification=\magstep1  \def\b{\bigskip}
\def\s{\smallskip} \def\c{\centerline}
  \c {\bf Electrino bubbles and relational entanglement }\b \c {Italo Vecchi} \c {Vicolo del Leoncorno 5 - 44100 Ferrara - Italy}  \c {email: vecchi@weirdtech.com}  \b

{\bf Abstract:} {\sl We argue that the phenomena exhibited by bubbles forming around free electrons in liquid helium and examined by Maris in his controversial 2000 paper point to the experimental relevance of relational entanglement. An experiment to verify/disprove the relevant argument is suggested.}
\b

In [1] Maris aims to "bring into very sharp focus some of the uncertainties of quantum measurement theory" and argues that "quantum mechanics does not make clear predictions for the results of
measurements on systems" like those he examines. The present note addresses the point raised by Maris, arguing that the experimantal results examined in [1] require  entanglement to be viewed in a relational setting, as a property of the  measurement process enforcing constraints on distinct measurement outcomes ([2],[3]). Furthermore, some critiques to Maris' model by other authors are briefly discussed. Our argument is directly related to Rovelli-Smerlak's recent work [3] about entanglement (see [2] for a similar argument in a "quantum coin" setting). The relevant relational argument may be summarised as follows. If we formulate the classical EPR problem in terms of observers Alice and Bob, interaction/measurement of Alice and the electron results, as long as  evolution remains unitary, in a superposition of "Alice measuring spin up" and "Alice measuring spin down". Entanglement appears then as a property of the local measurement/information-exchange between superposed Alice and Bob , when measurement outcomes are matched. In this setting entanglement nonlocality disappears, together with the hidden assumptions that spawned it. \s
According to our relational approach, if the state vector encodes  observer's knowledge about measurement outcomes, then  Bob's and Alice's subjective knowledge is encoded by two distinct state vectors. Measurement-induced changes in one of them will not affect the other as long as no information is exchanged locally between Bob and Alice. It is only when information is exchanged, i.e. when Bob and Alice measure each other's state locally, that entanglement comes into play, enforcing constraints on measurement outcomes on which Bob and Alice agree. That holds also when Bob and Alice are the same person or device, i.e. upon subsequent measurements the resulting superposed instances of the observer will be matched appropriately.  We contend that  regarding entanglement as a constraint on distinct measurement outcomes, rather than as an intrinsic property of the wave-function, is decisive for a correct interpretation of Maris' model. Measurement is a local process that affects the wave function only locally, while entanglement steers the outcome of subsequent measurements.\s 
In [1] a series of experiments are described in terms  bubbles arising around optically excited electrons in liquid helium. It is argued that, in order to minimise the  energy of the bubble resulting from optical excitation, the electrons split, giving rise to "electrino" bubbles, which contain fractional electrons. Such a model accounts for experimental results whose nature is controversial (cf. [4]). In [5] it is rightly argued that electrinos correspond to entangled electrons with fractional amplitudes. We contend here that the properties of the bubbles arising around such entangled electrons with fractional amplitudes require certain semantic models about particles and physical objects to be reexamined in the light of the understanding of entanglement which has been sketched and referenced above. In the sequel we will refer to the entangled electrons with fractional amplitudes as electrinos (cf. [1])  and we will discuss their properties.\s
The electron is a charge carrier. Each electrino carries the full electron charge, but gives rise to fractional amplitude bubbles carrying only a fraction of the energy of the bubble from which they arise through splitting. Now, let us recall that, as famously pointed out by Zeh in [8], "there are no particles". The notion of particle, as well as any notion of physical object, is just a handy shorthand for a cluster of observables and for the corresponding measurement outcomes . Entanglement is a property of the measurement process, since it relates different measurement outcomes ([7]). Strictly speaking, it is not the particles or the objects that are entangled, but the measurements relative to the observables that are associated with them. At the core of our argument is the distinction between different observables  associated with the particle or the object. Some of those measurements may be entangled while others are not.  In the case under examination, we argue that the electron charge measurements are entangled, while the bubble energy measurements are not. In other words and more concretely, upon measurement of the electron's charge at the detector, a fractional-amplitude bubble will  release photons locally, regardless of  whether the fractional-amplitude electron in it triggers detection or is found to be locally absent. Similarly, other bubbles entangled with the one being measured will release their photons locally only when they will be subjected to measurement, i.e. when the electron charge will be measured (i.e. detected or found to be locally absent). Such measurements will be found to be entangled so as to preserve conservation of charge and of energy, but they will not enable any superluminal signalling, due to their purely local nature.   On this basis we argue  that the implicit  and erroneous assumption that entanglement involves nonlocality  is at the core of Altschul and Rebbi's ([6]) objections against Maris' electrino model.\s
In [6] Maris' model is dismissed, while ignoring  the relevant experimental data, which are not even cited. Failure to account for the observed experimental results appears to be a common feature of the papers where Maris' model is dismissed (see [4] for references). In fact the model proposed in [6] contemplates "full sized bubbles" which are incompatible with the observed phenomena surveyed in [1]. The argument in [6] is actually based on a 1-dimensional model, which cannot capture the splitting induced by the minimization of the bubbles'surface energy. In [6] the authors claim that Maris' model would enable superluminal signalling. They state that "if the position of the electron is measured, the system will collapse into a state with only a single bubble, surrounding the location where the electron is found". On the basis of the argument sketched above, we argue that such statement in [6] is erroneous and that the underlying argument is flawed, based as it is on a misunderstanding of the nature of entanglement and on an inappropriate semantic model, i.e. on an inappropriate correspondence between different observables based on arbitrary semantic constructs. In simpler words and as argued above, upon measurement of the electron charge the bubbles that are not being directly measured will stay put.\s
An appropriate experiment, aimed at verifying that an electrino bubble releases photons also when the fractional amplitude electron that spawns it is not detected, should enable verification/falsification of the argument presented here. Such an experiment is actually foreshadowed in the remarks about measurement of "exotic ions" on page 201 of [1].\b
\c{ \bf References} \b 
[1] H.J. Maris "On the Fission of Elementary Particles and the Evidence for Fractional Electrons in Liquid Helium", Journal of Low Temperature Physics , 120, 3/4 (2000) pp. 173-204 \s
[2] I.Vecchi "Are classical probabilities instances of quantum amplitudes?", quant-ph/0206147\s
[3] C.Rovelli, M.Smerlak "Relational EPR", arXiv:quant-ph/0604064\s
[4] H.J. Maris "Properties of Electron Bubbles in Liquid Helium", Journal of Low Temperature Physics, Vol.132, Nos.1/2,July 2003\s
[5]  R. Jackiw, C. Rebbi, J.R. Schrieffer "Fractional Electrons in Liquid Helium?" , arXiv:cond-mat/0012370\s
[6] B. Altschul, C. Rebbi "Analysis of a Toy Model of Electron 'Splitting'", arXiv:cond-mat/0211096 \s
[7] I.Vecchi "Is entanglement observer-dependent?", arXiv:quant-ph/0106003\s
[8] H.D.Zeh "There are no Quantum Jumps, nor are there Particles!", Physics Letters A, 172 (1993) 189-192. \s
\end